\documentclass[superscriptaddress,prl,floatfix,twocolumn,amsmath,amssymb,aps]{revtex4-1}

\usepackage{graphicx}
\usepackage{amsmath}
\usepackage{amssymb}
\usepackage{dcolumn}
\usepackage{color}
\usepackage{multirow}
\usepackage{bm}
\usepackage{subfigure}

\begin{document}
\title{Dynamic phase transition theory}
\author{Qi-Jun Ye}
\affiliation{State Key Laboratory for Artificial Microstructure and Mesoscopic Physics, Frontier Science Center for Nano-optoelectronics and School of Physics, Peking University, Beijing 100871, P. R. China}
\author{Xin-Zheng Li}
\email{xzli@pku.edu.cn}
\affiliation{State Key Laboratory for Artificial Microstructure and Mesoscopic Physics, Frontier Science Center for Nano-optoelectronics and School of Physics, Peking University, Beijing 100871, P. R. China}
\affiliation{Interdisciplinary Institute of Light-Element Quantum Materials, Research Center for Light-Element Advanced Materials, and Collaborative Innovation Center of Quantum Matter, Peking University, Beijing 100871, P. R. China}
\affiliation{Peking University Yangtze Delta Institute of Optoelectronics, Nantong, Jiangsu 226010, P. R. China}
\date{\today}

\begin{abstract}
    Thermodynamic conventions suffer from describing dynamical distinctions, especially when the structural and energetic changes induced by rare events are insignificant.
    By using the ensemble theory in the trajectory space, we present a statistical approach to address this problem.
    Rather than spatial particle-particle interaction which dominates thermodynamics, the temporal correlation of events dominates the dynamics.
    The zeros of dynamic partition function mark phase transitions in the space-time, i.e., dynamic phase transition (DPT), as Yang and Lee formulate traditional phase transitions, and hence determine dynamic phases on both sides of the zeros.
    Analogous to the role of temperature (pressure) as thermal (mechanical) potential, we interpret the controlling variable of DPT, i.e., dynamic field, as the dynamical potential.
    These findings offer possibility towards a unified picture of phase and phase transition.
\end{abstract}

\maketitle
\begin{figure}[b]
    \centering
    \includegraphics[width=0.95\linewidth]{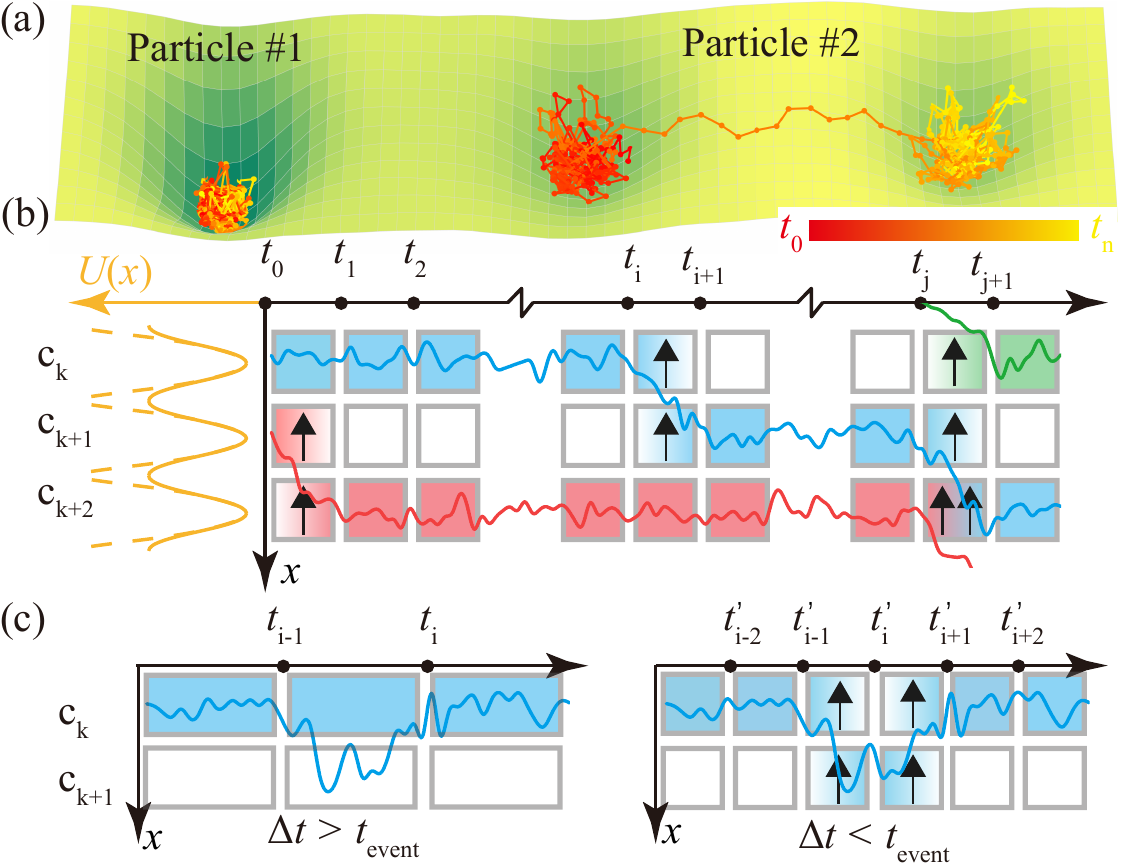}
    \caption{Description of dynamic states. (a) Schematic of realistic trajectories in the potential energy surface (PES). At moderate $T$s, particles would almost be localized at the minimals of the PES (like \#1 and \#2) while occasionally transfer to another (like \#2). (b) A coarse gained description of trajectories by discretization of the space-time. The components $c_k$s are defined according to the PES. The events (inter-component motions) are highlighted as up arrows while the localized motions are summarized. (c) The observation results for choosing different temporal interval $\Delta t$s. Fundamental dynamic information can only be contained by choosing $\Delta t$ less than the typical timescale of the event.}
    \label{fig1}
\end{figure}

Since the mid-1800s, the efforts to understand gas laws in a microscopic manner end up with the foundation of statistical physics.
The Boltzmann-Gibbs's description of equilibrium states is based on ``ensemble''~\cite{Gibbs2010}, i.e., all possible states ideally assemble to represent the real system.
The properties of the system can be inferred by averaging over an equilibrium ensemble of states, as
\begin{equation}
    \label{eq:exp-ensemble}
    \begin{split}
        \mathbb{E}(A) &=  \sum_{\text{all states}} p(i^{\text{th}}{\text{-state}}) A(i^{\text{th}}{\text{-state}})\\
        &=  \langle A \rangle = \sum_{\{\Omega\}} p(\{\Omega\}) A(\{\Omega\}),
    \end{split}
\end{equation}
where $A$ is an arbitrary observable, and the states are copies of the system classified by a set of thermodynamic extensive quantities $\{\Omega\}$, such as energy $E$ and number of particles $N$~\footnote{``A state with $T$=273~K'' sounds commonplace. But strictly speaking, only extensive quantities can be used to describe microstates, while the intensive quantities can be only defined macroscopically via many microstates.}.
The equilibrium possibility distribution function (PDF) follows the well-known exponential form
\begin{equation}
    \label{probability}
    p(\{\Omega\}) \sim \prod_{\{\Omega\}} \exp\left[-\frac{\Omega \cdot C_\Omega}{k_{\text{B}}}\right],
\end{equation}
where $C_\Omega$ is the conjugated potential such as thermal potential $T^{-1}$ and chemical potential $\mu T^{-1}$.
Phase transition marks when there are visible distinctions between two sets of uniformed states on extensive quantities and hence the measured observables, within the control of conjugated potential.
Despite the controversy on ergodicity~\cite{Palmer1982}, the above framework is in principle applicable for all types of systems and phase transitions therein.
Unfortunately, its power has been confined in thermodynamic conventions (TC), where the derived concepts like static equilibrium and thermodynamic phase transition (TPT) are inherently flawed when describing systems with internal flows and dynamic pathways, such as premelted solids~\cite{Dash2006,Martorell2013}, percolations~\cite{Grassberger1983,Li2015}, even complex networks in biology and economics~\cite{Mizuno2007,Pastor-Satorras2015}.
To fill this gap, the statistical theory must be reformulated on a dynamic basis~\cite{Palmer1984}.
Among pioneering practices, the space-time is poised to play the core role.
The dynamical distinctions become apparent if one examines trajectories of full spatial-temporal features, other than structural orders and thermal free energies according to TC.
In seeking the PDF over trajectories, Jaynes et al. introduced the principle of maximum caliber as the dynamic analogue of the principle of maximum entropy~\cite{Jaynes1957,Shore1980,Presse2013,Ghosh2020}.
Contrary to common views on ``glass transition'', Hedges et al. interpreted it as a phase transition in space-time by employing an artificial field~\cite{Hedges2009,Chandler2010}.
Outside these model systems, this idea has been realized recently in realistic condensed matters.
Ye et al. found that two states corresponding to ice VII and dynamic ice VII can only be discriminated rigorously from a dynamic perspective~\cite{Ye2021}.
However, the existence of well-defined dynamic phases and an inclusive framework interpreting corresponded phase transitions remain unclear.
It is imperative to seek a language which relates those novel dynamic phenomena to the familiar theory of equilibrium.
In this article, we present a statistical ensemble approach to describe the spatial-temporal equilibrium states with dynamical distinctions, namely the dynamic phase transition (DPT) theory.
All possible trajectories within unified dynamic field assemble to represent a single equilibrium dynamic state (EDS) of real system.
We found the dynamic field, a quantity formerly used to artificially access different EDSs, is intrinsic to the EDS and can be determined in a microscopic manner.
With trajectories classified into series of coarse-grained events, the event-event correlation dominates the dynamics, just like the Hamiltonian of particle-particle interaction dominates the thermodynamics.
LY zeros of the dynamic partition function are used to confirm the existence of phases (sets of EDSs uniformed in dynamics) and phase transitions therein, where the dynamic field plays the role of controlling variable like $T$ in TPT.
We illustrate these concepts for high pressure~($P$) ice, however, the underlying ideas can be generalized to broad complex systems.
While thermal distinctions are always accompanied by dynamical ones, the converse does not hold.
We are interested in systems where dynamical distinctions emerge without experiencing TPT and are hence invisible in TC.
Thus, we examine the trajectories, the fundamental elements for dynamics.
A single trajectory is represented, as
\begin{equation}
    \label{phase point}
    \mathcal{P}_{\text{beg}\to \text{end}}[x(t)]=\prod_{j}\delta (x(t_j)-x_j),
\end{equation}with $x(t)=\{\cdots,q_i(t),\cdots,p_i(t),\cdots\}$ the spatial coordinates, and $t_j=t_{\text{beg}}+j\Delta t~(j=0,\cdots,N_t)$ the discretized temporal coordinates.
Eq.~(\ref{phase point}) thermalizes to Gibbs's phase space when the temporal degrees of freedom (DOF) is trivial: points of the phase space are summed ignoring their temporal ordering, e.g., in equilibrium solids where the particles are simply confined to a region throughout the observation timescale (Fig.~\ref{fig1}(a)).
Palmer called such a region the ``component"~\cite{Palmer1982}, and  intra-component motions are captured as fluctuations in TC.
However, the temporal DOF becomes nontrivial when $T$ is moderate: there are rare but significant inter-component motions for particles moving farther away beyond thermal fluctuations~\cite{Geng2017,Komatsu2020,Wang2021}.
We note that it implied no TPT, since the distinctions in transport properties such as the diffusion coefficients and ionic mobilities arise within consistent structural orders and thermal free energies~\cite{Queyroux2020,Ye2021}.
The system is intuitively in a ``dynamical equilibrium", since the dynamics of inter-component motions becomes homogeneous towards the long time limit~\cite{Ye2021}.
Meanwhile, by extending the ensemble theory to trajectory space, this idea can have a rigorous basis as show below.
The ensemble theory is powerful to make statistical inferences on macroscopic quantities, by averaging over a great number of independent replicas of system as microstates.
For specificity of the dynamic ensemble, two issues need to be resolved: how to describe dynamical microstates and how they distribute in the probability space.
The straightforward choice to describe dynamical microstates is by trajectories formulated in Eq.~(\ref{phase point}).
However, the original trajectories are so noisy that the dynamic features are overwhelmed by the predominant intra-component motions.
To highlight the inter-component motions, a coarse grain manner is applied by extracting only the timestamps where events occur.
Here, ``event'' is used to nominate the process that a single particle transfers to the nearest neighboring components, whose different types reveal the spatial connection between components.
Without loss of generality, we consider systems consisting of identical particles, where the overall properties can be inferred from the behavior of individual particle subjected to effective potentials.
This allows easier but equivalent observation from the trajectories of single particle rather than the system one.
On above bases, each independent dynamic microstate is finally expressed by single particle trajectory $\mathcal{T}$, as
\begin{equation}
    \label{trajectory}
    \mathcal{P}_{\text{beg}\to \text{end}}[x(t)]\sim\mathcal{T}(K)=\{\tau_1,\cdots,\tau_K\},
\end{equation}
where $K$ is the total number of events, and $\tau_k$ is the timestamp where $k$-th event occurs.

\begin{figure}[h]
    \centering
    \includegraphics[width=0.95\linewidth]{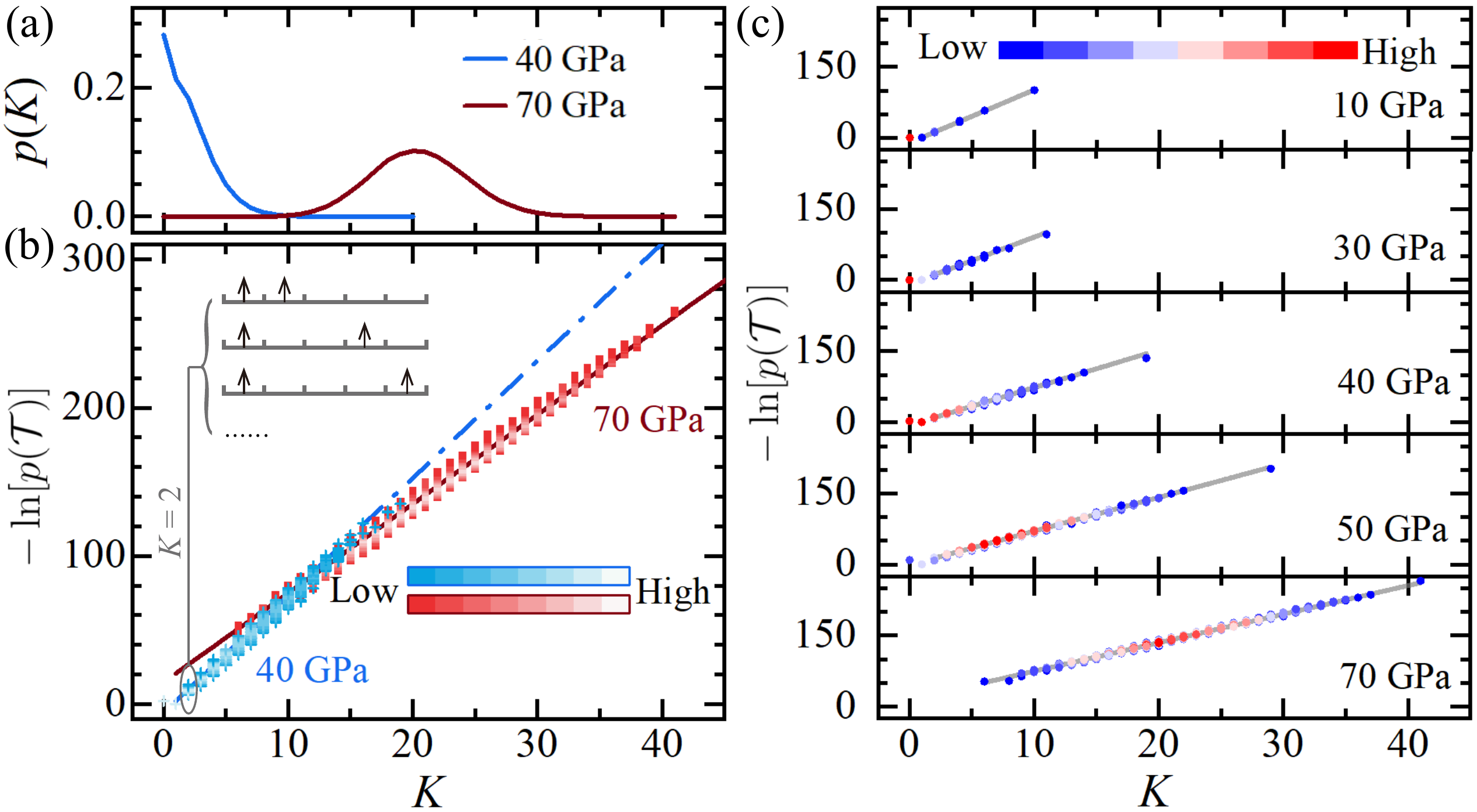}
    \caption{Perspectives macroscopically from $K$ and microscopically from $\mathcal{T}$. (a) The cumulated probability $p(K)$ according to macroscopic manner. Turning to a microscopic manner, i.e., emphasizing on each single $\mathcal{T}$, (b)(c) shows density plot of the effective free energy $-\ln [p(\mathcal{T})]$. The plots are enlarged in (b) to show details with $P=$~40~GPa (blue open squares) and 70~GPa (red solid dots). The inset of (b) show an example of looking into $K=2$ case, which consists of many different $\mathcal{T}$s. The solid and dashed lines in (b)(c) show the linear fitting results of $\mathcal{T}(K\neq0)$.  Each counted trajectory is in a timescale of $t=2~\text{ps}$.}
    \label{fig2}
\end{figure}

Then, we try to reveal the PDF of those dynamic microstates.
By making analogy to Eq.~(\ref{probability}), the PDF should be something like $p(K)\sim \exp\{-K\cdot C_K\}$.
With $K$ naturally establishing the dynamic extensive quantities, $C_K$ becomes the conjugated dynamic potential: when $C_K$ is increased (decreased), the events are suppressed (enhanced).
In fact, Hedges et al. firstly proposed dynamic field $s$ to be $C_K$ but only employ it as an artificial tool to access different dynamic pathways~\cite{Hedges2009}.
We adopt this notation.
In the following, we demonstrate $s$ is intrinsic to the dynamic ensemble and obtain PDF upon the insights into microscopic correlations.
It means $s$ can be rigorously determined from statistics on $\mathcal{T}(K)$, just like other thermal intensive quantities, e.g., $T$ and $P$.
Dividing the space-time into pieces, as shown in Fig.~\ref{fig1}(b), Eq.~(\ref{trajectory}) implies a strong connection with lattice gas (LG) model~\cite{Lee1952,Fisher1967}.
The occurrence of events in spatial-temporal pieces is analogous to the occupation of particles on lattice sites.
While the spatial DOF has been handled by Hamiltonian of the system in TC, the temporal DOF is underappreciated.
We focus on the latter and note an interacting potential for the temporal DOF can be obtained by formulating an LG-liked grand partition function, as
\begin{equation}
    \label{hamiltonian}
    \begin{split}
        Z(s) &= \sum_{n_i=0,1} \exp \left\{-k_{\text{D}}^{-1}\left[s\sum_{i=1}^{N_t} n_i+\sum_{\left(i,i^\prime\right)}^{N_t} \phi_{ii^\prime}n_i n_{i^\prime}\right]\right\}\\
        &=\sum_{\mathcal{T}(K)} \exp \left\{-\frac{sK+U_0(\mathcal{T})}{k_{\text{D}}}\right\}.
    \end{split}
\end{equation}
The first line manifests a trivial LG model except for $n_i=n(t_i)=0,1$ being replaced from the occupations of lattice sites to the occurrence of events at $t_i$, and $\phi_{ii^\prime}$ specifies the interacting coefficients.
Notice that $\sum_{i=1}^{N_t} n_i=\sum_{i=1}^{N_t}\sum_{j=1}^K \delta(t_i-\tau_j)=K$, it comes to the final form.
Here, $U_0(\mathcal{T})$ is the internal temporal interaction of $\mathcal{T}$ \footnote{The form of the interacting potential might be more complex than the pairwise one. We adopt the pairwise for convenience of illustration and later computation, but the treatments are analogous for other forms of interacting potential.} and $s$ acts on $\mathcal{T}$ as an external field (in LG it is the chemical potential).
$U_0(\mathcal{T})$ for $\mathcal{T}$s and $s$ are the analogues to particle-particle interaction $U(x)$ for spatial configuration $x$ and the chemical potential in LG model, respectively.
Almost all events can be tracked by choosing $\Delta t$ less than the typical timescale of events (Fig. \ref{fig1}(c)).
The meaning of the coefficient $k_{\text{D}}$ with D for the dynamics shall be discussed later.
Now we can derive the PDF of dynamic microstates $\mathcal{T}$s.
As $Z(s)$ in Eq.~(\ref{hamiltonian}) contains all important dynamic information, we call it the dynamic partition function.
The PDF is given by $p(\mathcal{T})=Z(s)^{-1} \exp\{-k_D^{-1}(sK+U_0(\mathcal{T}))\}$.
This relation holds for arbitrary $\mathcal{T}$ in the ensemble, which immediately leads to an expression for $s$, as
\begin{equation}
    \label{micro definition of s}
    s = -\lim_{K\to\infty}\left[k_{\text{D}} \ln p(\mathcal{T}) - U_0(\mathcal{T})\right]/K + \text{const.},
\end{equation}
where $K\to\infty$ is equivalent to the long time limit $t\to\infty$.
Eq.~(\ref{micro definition of s}) defines the dynamic field microscopically, while in previous studies it is either artificially assigned~\cite{Hedges2009} or derived macroscopically~\cite{Ye2021}.
More importantly, it clarifies the meaning of the EDS: the set of dynamic microstates conforming to a unique $s$.
Standing on the perceptions of ``static equilibrium'', events can be easily attributed to that system is driven out of equilibrium.
However, in the sense of ``dynamic equilibrium'', they may belong to an EDS once the sampled trajectories give converged $s$.
This welcomes some thought-to-be nonequilibrium phenomena exhibiting steady dynamic profiles into a unified statistical framework with traditional equilibrium phenomena.
With events being the subject, it is more contextual to interpret $U_0(\mathcal{T})$ as the correlations.
A fundamental question emerges: how to connect the microscopic $U_0(\mathcal{T})$ with macroscopic energy.
Especially when thermal microstates are considered simultaneously, e.g. in the constant $T$ and $s$ ensemble, the total partition function should contain both thermal and dynamic part, as
\begin{equation}
    Z(s,T) = \sum_{\Omega(E)}\sum_{\mathcal{T}(K)} \exp \left\{-\frac{T^{-1}\cdot E + S}{k_{\text{B}}}-\frac{s\cdot K + U_0(\mathcal{T})}{k_{\text{D}}}\right\}.
\end{equation}
The dynamic terms contribute to the thermal free energy effectively, with a ratio of $k_{\text{D}}/k_{\text{B}}$.
While $k_{\text{B}}$ bridges energy to typical microstates in spatial DOF, a new constant $k_{\text{D}}$ is intrinsic to perform the same role for dynamic microstates in space-time.
Considering the fact that EDSs are often inconspicuous from the energetic perspective, $k_{\text{D}}$ should be several magnitudes less than $k_{\text{B}}$.
Fortunately, this does not hinder the exploration of EDSs when only dynamics are nontrivial, since $k_{\text{D}}$ can be extracted into the units of $s$ and $U_0$ without specifying its value.
We use this notation in the following.
The remaining is to validate these concepts in realistic condensed matter.
A straightforward evidence is the existence of DPT invisible in TC but visible in our theory.
We choose high-$P$ bcc ice as the example since it's a typical system where different dynamic states emerge from the nature of rare events.
At 500~K and 10-70~GPa, the diffusion coefficients of protons change from solid-like to liquid-like implying DPT between inactive and active phases, whilst the structure order and thermal free energies remain consistent.
Another reason is its concise dynamic features that the robust oxygen skeleton offers well-defined components and constrains above transfers to be the same type.
To sample the dynamic microstates $\mathcal{T}$s and derive $p(\mathcal{T})$, we performed a large amount of molecular dynamic (MD) simulations, as detailed in Ref.~\cite{Ye2021,SI}.
We shall gain new insights by digging into trajectories.
Only information on macroscopic dynamic properties are effectively utilized in previous studies.
For example, the events are counted according to its number of occurrences, as the cumulated probability $p(K_0)=\sum _{\mathcal{T}(K=K_0)}p(\mathcal{T}(K))$ according to $K$ partition shown by Fig.~\ref{fig2}(a).
There is nothing tricky, but their intrinsic commons are not obvious till being reviewed in a microscopic manner.
We decompose the distribution to the level of single $\mathcal{T}$, i.e., using $p(\mathcal{T})$ instead of $p(K)$, as the density plot of effective free energy $-\ln[p(\mathcal{T})]$ in Fig.~\ref{fig2}(b).
All cases are consistent: the $\mathcal{T}(K>0)$s just lay aside a linear line, with varied slopes for different $P$s (Fig.~\ref{fig2}(c)).
We shall resort to $s$ and $U_0(\mathcal{T})$ to explain this.

Systems with uniformed structural order are expected to exhibit uniformed $U_0$.
In fact, this is a hypothesis on reductionism by considering events as the fundamental elements.
Within one structure, different external conditions such as $(T,P)$s exert similar dynamic constraints but with different intensities $s$s.
Based on the fact that system has converged macroscopic dynamic properties, we consider the system controlled by $(T,P)$ in thermodynamics is also controlled by a single $s$ in dynamics.
The consistent linearity in Fig.~\ref{fig2}(c) exactly reflects the uniformed $U_0$ and existence of EDSs, and hence we believe the hypothesis is true.
Note that $U_0$ and $s$ can be solved simultaneously via the equation series transformed from Eq.~(\ref{micro definition of s}), as
\begin{equation}
    \label{deriving dynamic field}
    U_0(\mathcal{T}(s, (T,P))) + s\cdot K = -k_{\text{D}} \ln p\left[\mathcal{T}(s, (T,P)) \right],
\end{equation}
where the realistic $\mathcal{T}(s,(T,P))$s have both thermal and dynamic dependencies.
In practice, the $\mathcal{T}(s,(T,P))$s are sampled from $NVE$ MD simulations to eliminate the effects of artificial bath, where initial configurations are randomly picked from precedent $NPT/NVT$ MD simulations to represent the environments $(s,(T,P))$s.

\begin{figure}[b]
    \centering
    \includegraphics[width=0.95\linewidth]{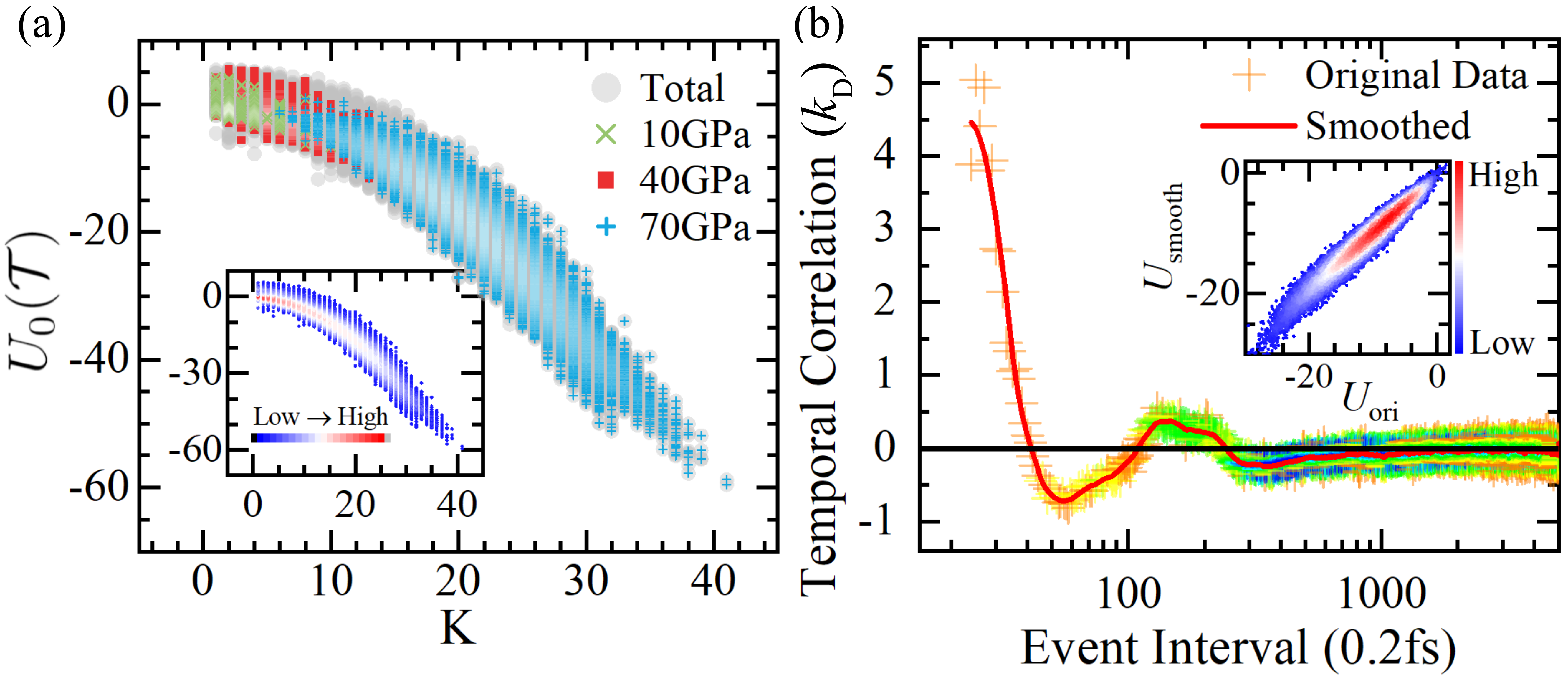}
    \caption{The uniformed $U_0(\mathcal{T})$ underlying EDSs. (a) Density plot of $U_0(\mathcal{T})$ from $\mathcal{T}$s at 10, 40, 70~GPa, and the total are shown in the inset. (b) The event-event interaction solved from Eq.~(\ref{deriving dynamic field}). Inset of (b) show the comparison between the smoothed potential (red line in (b)) and original solved one.}
    \label{fig3}
\end{figure}

\begin{figure}[t]
    \centering
    \includegraphics[width=0.95\linewidth]{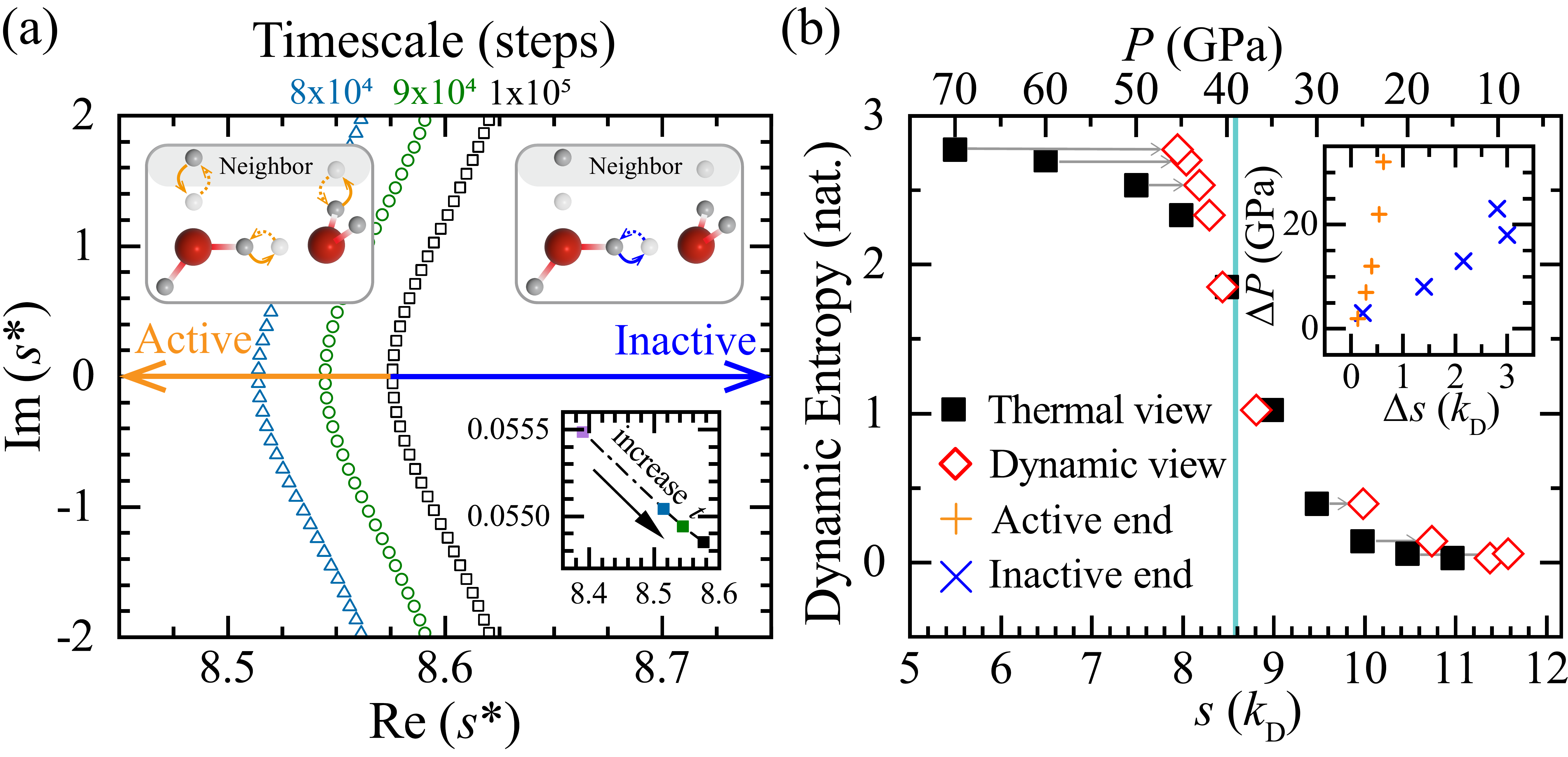}
    \caption{Evidences for dynamic phase transition. (a) The temporal evolution of complex zeros of dynamic partition function. Bottom inset of (a): the tendency of dropping to real axis of the nearest zeros. The results are shown with timescale from $8\times 10^4$ to $1\times 10^5$ steps and stepsize of 0.2~fs. For realistic systems, $Z(s)$ can only take values of $s$ on the real axis. The zeros define phases on both sides: the inactive end with single transfer, and the active end with cooperated transfers. Schematic of their different behaviors are shown in the top insets of (a), where oxygen atoms, protons, and transferred protons are in red, gray, and light gray, respectively. (b) The changes of dynamic entropy in different perspectives from flat scale of thermal (solid black squares) and dynamic (open red diamonds) variables. The transition point $P_c$ and $s_c$ for both views has been aligned coincident as the canyon vertical line. Inset of (b) shows the distinct $\Delta P/ \Delta s$ behaviors on both sides of the transition point.}
    \label{fig4}
\end{figure}

In Ref.~\cite{Ye2021}, we employ a physical intuition as the iso-structural ensemble to explain above internal consistency.
Here in Fig.~\ref{fig3}(a), the distribution of $U_0$ according to $\mathcal{T}$s provides its basis: $\mathcal{T}$s at different $P$s exactly contribute to different zones of the uniformed total one.
A reasonable $U_0$ is expected to be infinite repulsive at small event interval and negatively converged to zero, alike the particle-particle interaction in LG.
The solved $U_0$ in Fig.~\ref{fig3}(b) is consistent with these theoretical expectations.
Besides, the repulsive peak near 100$\sim$200 steps shows significant divergence from the regular interaction models such as the Lennard-Jones type, and $U_0$ is of long range though weak magnitude.
In the meantime, $s$ can be obtained providing the norm $s=0$ when $t\to 0$, which is applied to eliminate the constant term in Eq.~(\ref{micro definition of s}).
The change of $s$-dependency on $P$ is witnessed around $P=39$~GPa, exactly where the dynamic phase transition is reported~\cite{Ye2021}.
Please see SI for calculation and discussion on such system-specific features~\cite{SI}.
From the identification to LG, dynamic phase transition can be established.
Eq.~(\ref{hamiltonian}) can be rewritten with the argument $\gamma = e^{-s/k_{\text{D}}}$, as
\begin{equation}
    \label{zeros}
    Z(s)=Z(\gamma)=\sum_{\mathcal{T}} e^{-k_{\text{D}}^{-1}U_0(\mathcal{T})} \gamma^K = \sum_{K} p(K) \gamma^K.
\end{equation}
Note that the dynamic partition function is a finite polynomial of $\gamma$ since $K$ can only take positive integers.
Following Yang et al.~\cite{Yang1952}, we can factorize Eq.~(\ref{zeros}) by its complex zeros, as
\begin{equation}
    Z(\gamma)=p(K=0)\prod_{i=0}^{K_{\text{max}}} \left(1-\frac{\gamma}{\gamma_i}\right),
\end{equation}
where $\gamma_i$ are the roots of $Z(\gamma) = 0$.
These roots are called Lee-Yang zeros~\cite{Yang1952} or Fisher zeros~\cite{Fisher1965}.
Yang et al. proved that if these zeros do close in onto the positive real axis, then thermodynamic properties would experience abrupt changes when driving system through the zeros, i.e., a phase transition occurs.
We note all the assumptions to derive this are fulfilled in dynamics, with replacing volume and the number of atoms to timescale and the number of events, respectively.
As shown in Fig.~\ref{fig4}(a), the numerical results towards the long-time limit confirm that the aforementioned transition is well-defined phase transition.
Since $U_0(\mathcal{T})$ exhibits none singularity, it is fascinating that nonanalytic behavior of $Z(s)$ emerges when approaching the long-time limit.
The knowledge of inactive end cannot predict the cooperated transfer in the active end, and vice versa.

While acknowledging the fact that there are two phases, only a slow transition can be perceived from conventional thermal view in Fig.~\ref{fig4}(b).
This poor tendency cannot be improved by increasing the scale of system, which seemingly disobey the phase transition theory.
Employing the dynamic view, however, the two sides show discriminated $s$ dependencies.
The configurations of $P>P_c$ shrink to a narrow $s$ region, while the $P<P_c$ ones remain in a loosed $s$ region.
It reveals that the variation of dynamic behaviors is not homogeneous in the thermal view.
This discrepancy is more clear by $\Delta s$-$\Delta P$ curves in the inset of Fig.~\ref{fig4}(b).
The success or failure is evident from the facts that $s$ is the intrinsic dynamic intensive quantity while $P$ is not.
To summarize, we establish a link between microscopic trajectories and macroscopic dynamics.
The framework inherited from thermodynamics is shown qualified to treat dynamic phenomena, providing
a unified statistical picture on phase and phase transition.
This theory extends the studies of state space from static profiles to the rule of dynamic evolution, whilst the concepts such as $s$ and $U(\mathcal{T})$ build the appreciation for the nature of dynamics.
Recent pivotal insights stimulated from the temporal DOF, such as the time crystal emerged from the breaking of temporal translational symmetry~\cite{Wilczek2012,Shapere2012,Zhang2017}, and quantum dynamical phase transition (QDPT) manifesting temporal criticality~\cite{Heyl2013,Heyl2018,Smith2016}, may also benefit.
EDS reveal the patterns behind noisy trajectories and stochastic rare events in complex systems, implying the existence of quasi time crystal.
Sharing the same challenge of identifying states beyond equilibrium paradigm,
QDPT utilizes Loschmidt amplitude rather than the conventional energetic quantities~\cite{Heyl2018}.
The coefficient $k_{\text{D}}$ might be the key to realize the free energy analogue in QDPT, promoting the urgency of making correspondence of these theories.
%
%
\begin{acknowledgments}
    We acknowledge helpful discussions with H. T. Quan, L. Zhuang, Y. C. Zhu, F. C. Wang, J. X. Zeng, and H. Y. Yang.
    We are supported by the National Basic Research Programs of China under Grand No. 2021YFA1400503,
    the National Science Foundation of China under Grant No. 11934003, the Beijing Natural Science Foundation under Grant No. Z200004, and the Strategic Priority Research Program of the Chinese Academy of Sciences Grant No. XDB33010400.
    The computational resources were provided by the supercomputer center in Peking University, China.
\end{acknowledgments}


\begin{thebibliography}{40}%
    \makeatletter
    \providecommand \@ifxundefined [1]{%
        \@ifx{#1\undefined}
    }%
    \providecommand \@ifnum [1]{%
        \ifnum #1\expandafter \@firstoftwo
        \else \expandafter \@secondoftwo
        \fi
    }%
    \providecommand \@ifx [1]{%
        \ifx #1\expandafter \@firstoftwo
        \else \expandafter \@secondoftwo
        \fi
    }%
    \providecommand \natexlab [1]{#1}%
    \providecommand \enquote  [1]{``#1''}%
    \providecommand \bibnamefont  [1]{#1}%
    \providecommand \bibfnamefont [1]{#1}%
    \providecommand \citenamefont [1]{#1}%
    \providecommand \href@noop [0]{\@secondoftwo}%
    \providecommand \href [0]{\begingroup \@sanitize@url \@href}%
    \providecommand \@href[1]{\@@startlink{#1}\@@href}%
    \providecommand \@@href[1]{\endgroup#1\@@endlink}%
    \providecommand \@sanitize@url [0]{\catcode `\\12\catcode `\$12\catcode
        `\&12\catcode `\#12\catcode `\^12\catcode `\_12\catcode `\%12\relax}%
    \providecommand \@@startlink[1]{}%
    \providecommand \@@endlink[0]{}%
    \providecommand \url  [0]{\begingroup\@sanitize@url \@url }%
    \providecommand \@url [1]{\endgroup\@href {#1}{\urlprefix }}%
    \providecommand \urlprefix  [0]{URL }%
    \providecommand \Eprint [0]{\href }%
    \providecommand \doibase [0]{http://dx.doi.org/}%
    \providecommand \selectlanguage [0]{\@gobble}%
    \providecommand \bibinfo  [0]{\@secondoftwo}%
    \providecommand \bibfield  [0]{\@secondoftwo}%
    \providecommand \translation [1]{[#1]}%
    \providecommand \BibitemOpen [0]{}%
    \providecommand \bibitemStop [0]{}%
    \providecommand \bibitemNoStop [0]{.\EOS\space}%
    \providecommand \EOS [0]{\spacefactor3000\relax}%
    \providecommand \BibitemShut  [1]{\csname bibitem#1\endcsname}%
    \let\auto@bib@innerbib\@empty
    \bibitem [{\citenamefont {Gibbs}(2010)}]{Gibbs2010}%
    \BibitemOpen
    \bibfield  {author} {\bibinfo {author} {\bibfnamefont {J.~W.}\ \bibnamefont
            {Gibbs}},\ }\href {\doibase 10.1017/CBO9780511686948} {\emph {\bibinfo
            {title} {Elementary principles in statistical mechanics}}}\ (\bibinfo
    {publisher} {Cambridge University Press},\ \bibinfo {address} {Cambridge},\
    \bibinfo {year} {2010})\BibitemShut {NoStop}%
    \bibitem [{Note1()}]{Note1}%
    \BibitemOpen
    \bibinfo {note} {``A state with $T$=273~K'' sounds commonplace. But strictly
        speaking, only extensive quantities can be used to describe microstates,
        while the intensive quantities can be only defined macroscopically via many
        microstates.}\BibitemShut {Stop}%
    \bibitem [{\citenamefont {Palmer}(1982)}]{Palmer1982}%
    \BibitemOpen
    \bibfield  {author} {\bibinfo {author} {\bibfnamefont {R.~G.}\ \bibnamefont
            {Palmer}},\ }\href {\doibase 10.1080/00018738200101438} {\bibfield  {journal}
        {\bibinfo  {journal} {Adv. Phys.}\ }\textbf {\bibinfo {volume} {31}},\
        \bibinfo {pages} {669} (\bibinfo {year} {1982})}\BibitemShut {NoStop}%
    \bibitem [{\citenamefont {Dash}\ \emph {et~al.}(2006)\citenamefont {Dash},
                \citenamefont {Rempel},\ and\ \citenamefont {Wettlaufer}}]{Dash2006}%
    \BibitemOpen
    \bibfield  {author} {\bibinfo {author} {\bibfnamefont {J.~G.}\ \bibnamefont
            {Dash}}, \bibinfo {author} {\bibfnamefont {A.~W.}\ \bibnamefont {Rempel}}, \
        and\ \bibinfo {author} {\bibfnamefont {J.~S.}\ \bibnamefont {Wettlaufer}},\
    }\href {\doibase 10.1103/RevModPhys.78.695} {\bibfield  {journal} {\bibinfo
            {journal} {Rev. Mod. Phys.}\ }\textbf {\bibinfo {volume} {78}},\ \bibinfo
        {pages} {695} (\bibinfo {year} {2006})}\BibitemShut {NoStop}%
    \bibitem [{\citenamefont {Martorell}\ \emph {et~al.}(2013)\citenamefont
                {Martorell}, \citenamefont {Vocadlo}, \citenamefont {Brodholt},\ and\
                \citenamefont {Wood}}]{Martorell2013}%
    \BibitemOpen
    \bibfield  {author} {\bibinfo {author} {\bibfnamefont {B.}~\bibnamefont
            {Martorell}}, \bibinfo {author} {\bibfnamefont {L.}~\bibnamefont {Vocadlo}},
        \bibinfo {author} {\bibfnamefont {J.}~\bibnamefont {Brodholt}}, \ and\
        \bibinfo {author} {\bibfnamefont {I.~G.}\ \bibnamefont {Wood}},\ }\href
    {\doibase 10.1126/science.1243651} {\bibfield  {journal} {\bibinfo  {journal}
            {Science}\ }\textbf {\bibinfo {volume} {342}},\ \bibinfo {pages} {466}
        (\bibinfo {year} {2013})}\BibitemShut {NoStop}%
    \bibitem [{\citenamefont {Grassberger}(1983)}]{Grassberger1983}%
    \BibitemOpen
    \bibfield  {author} {\bibinfo {author} {\bibfnamefont {P.}~\bibnamefont
            {Grassberger}},\ }\href {\doibase 10.1016/0025-5564(82)90036-0} {\bibfield
        {journal} {\bibinfo  {journal} {Math. Biosci.}\ }\textbf {\bibinfo {volume}
            {63}},\ \bibinfo {pages} {157} (\bibinfo {year} {1983})}\BibitemShut
    {NoStop}%
    \bibitem [{\citenamefont {Li}\ \emph {et~al.}(2015)\citenamefont {Li},
                \citenamefont {Fu}, \citenamefont {Wang}, \citenamefont {Lu}, \citenamefont
                {Berezin}, \citenamefont {Stanley},\ and\ \citenamefont {Havlin}}]{Li2015}%
    \BibitemOpen
    \bibfield  {author} {\bibinfo {author} {\bibfnamefont {D.}~\bibnamefont
            {Li}}, \bibinfo {author} {\bibfnamefont {B.}~\bibnamefont {Fu}}, \bibinfo
        {author} {\bibfnamefont {Y.}~\bibnamefont {Wang}}, \bibinfo {author}
        {\bibfnamefont {G.}~\bibnamefont {Lu}}, \bibinfo {author} {\bibfnamefont
            {Y.}~\bibnamefont {Berezin}}, \bibinfo {author} {\bibfnamefont {H.~E.}\
            \bibnamefont {Stanley}}, \ and\ \bibinfo {author} {\bibfnamefont
            {S.}~\bibnamefont {Havlin}},\ }\href {\doibase 10.1073/pnas.1419185112}
    {\bibfield  {journal} {\bibinfo  {journal} {Proc. Natl. Acad. Sci. U. S. A.}\
        }\textbf {\bibinfo {volume} {112}},\ \bibinfo {pages} {669} (\bibinfo {year}
        {2015})}\BibitemShut {NoStop}%
    \bibitem [{\citenamefont {Mizuno}\ \emph {et~al.}(2007)\citenamefont {Mizuno},
                \citenamefont {Tardin}, \citenamefont {Schmidt},\ and\ \citenamefont
                {MacKintosh}}]{Mizuno2007}%
    \BibitemOpen
    \bibfield  {author} {\bibinfo {author} {\bibfnamefont {D.}~\bibnamefont
            {Mizuno}}, \bibinfo {author} {\bibfnamefont {C.}~\bibnamefont {Tardin}},
        \bibinfo {author} {\bibfnamefont {C.~F.}\ \bibnamefont {Schmidt}}, \ and\
        \bibinfo {author} {\bibfnamefont {F.~C.}\ \bibnamefont {MacKintosh}},\ }\href
    {\doibase 10.1126/science.1134404} {\bibfield  {journal} {\bibinfo  {journal}
            {Science}\ }\textbf {\bibinfo {volume} {315}},\ \bibinfo {pages} {370}
        (\bibinfo {year} {2007})}\BibitemShut {NoStop}%
    \bibitem [{\citenamefont {Pastor-Satorras}\ \emph {et~al.}(2015)\citenamefont
                {Pastor-Satorras}, \citenamefont {Castellano}, \citenamefont {{Van
                            Mieghem}},\ and\ \citenamefont {Vespignani}}]{Pastor-Satorras2015}%
    \BibitemOpen
    \bibfield  {author} {\bibinfo {author} {\bibfnamefont {R.}~\bibnamefont
            {Pastor-Satorras}}, \bibinfo {author} {\bibfnamefont {C.}~\bibnamefont
            {Castellano}}, \bibinfo {author} {\bibfnamefont {P.}~\bibnamefont {{Van
                        Mieghem}}}, \ and\ \bibinfo {author} {\bibfnamefont {A.}~\bibnamefont
            {Vespignani}},\ }\href {\doibase 10.1103/RevModPhys.87.925} {\bibfield
        {journal} {\bibinfo  {journal} {Rev. Mod. Phys.}\ }\textbf {\bibinfo {volume}
            {87}},\ \bibinfo {pages} {925} (\bibinfo {year} {2015})}\BibitemShut
    {NoStop}%
    \bibitem [{\citenamefont {Palmer}\ \emph {et~al.}(1984)\citenamefont {Palmer},
                \citenamefont {Stein}, \citenamefont {Abrahams},\ and\ \citenamefont
                {Anderson}}]{Palmer1984}%
    \BibitemOpen
    \bibfield  {author} {\bibinfo {author} {\bibfnamefont {R.~G.}\ \bibnamefont
            {Palmer}}, \bibinfo {author} {\bibfnamefont {D.~L.}\ \bibnamefont {Stein}},
        \bibinfo {author} {\bibfnamefont {E.}~\bibnamefont {Abrahams}}, \ and\
        \bibinfo {author} {\bibfnamefont {P.~W.}\ \bibnamefont {Anderson}},\ }\href
    {\doibase 10.1103/PhysRevLett.53.958} {\bibfield  {journal} {\bibinfo
            {journal} {Phys. Rev. Lett.}\ }\textbf {\bibinfo {volume} {53}},\ \bibinfo
        {pages} {958} (\bibinfo {year} {1984})}\BibitemShut {NoStop}%
    \bibitem [{\citenamefont {Jaynes}(1957)}]{Jaynes1957}%
    \BibitemOpen
    \bibfield  {author} {\bibinfo {author} {\bibfnamefont {E.~T.}\ \bibnamefont
            {Jaynes}},\ }\href {\doibase 10.1103/PhysRev.106.620} {\bibfield  {journal}
        {\bibinfo  {journal} {Phys. Rev.}\ }\textbf {\bibinfo {volume} {106}},\
        \bibinfo {pages} {620} (\bibinfo {year} {1957})}\BibitemShut {NoStop}%
    \bibitem [{\citenamefont {Shore}\ and\ \citenamefont
                {Johnson}(1980)}]{Shore1980}%
    \BibitemOpen
    \bibfield  {author} {\bibinfo {author} {\bibfnamefont {J.}~\bibnamefont
            {Shore}}\ and\ \bibinfo {author} {\bibfnamefont {R.}~\bibnamefont
            {Johnson}},\ }\href {\doibase 10.1109/TIT.1980.1056144} {\bibfield  {journal}
        {\bibinfo  {journal} {IEEE Trans. Inf. Theory}\ }\textbf {\bibinfo {volume}
            {26}},\ \bibinfo {pages} {26} (\bibinfo {year} {1980})}\BibitemShut {NoStop}%
    \bibitem [{\citenamefont {Press{\'{e}}}\ \emph {et~al.}(2013)\citenamefont
    {Press{\'{e}}}, \citenamefont {Ghosh}, \citenamefont {Lee},\ and\
    \citenamefont {Dill}}]{Presse2013}%
    \BibitemOpen
    \bibfield  {author} {\bibinfo {author} {\bibfnamefont {S.}~\bibnamefont
    {Press{\'{e}}}}, \bibinfo {author} {\bibfnamefont {K.}~\bibnamefont {Ghosh}},
    \bibinfo {author} {\bibfnamefont {J.}~\bibnamefont {Lee}}, \ and\ \bibinfo
    {author} {\bibfnamefont {K.~A.}\ \bibnamefont {Dill}},\ }\href {\doibase
        10.1103/RevModPhys.85.1115} {\bibfield  {journal} {\bibinfo  {journal} {Rev.
                Mod. Phys.}\ }\textbf {\bibinfo {volume} {85}},\ \bibinfo {pages} {1115}
        (\bibinfo {year} {2013})}\BibitemShut {NoStop}%
    \bibitem [{\citenamefont {Ghosh}\ \emph {et~al.}(2020)\citenamefont {Ghosh},
                \citenamefont {Dixit}, \citenamefont {Agozzino},\ and\ \citenamefont
                {Dill}}]{Ghosh2020}%
    \BibitemOpen
    \bibfield  {author} {\bibinfo {author} {\bibfnamefont {K.}~\bibnamefont
            {Ghosh}}, \bibinfo {author} {\bibfnamefont {P.~D.}\ \bibnamefont {Dixit}},
        \bibinfo {author} {\bibfnamefont {L.}~\bibnamefont {Agozzino}}, \ and\
        \bibinfo {author} {\bibfnamefont {K.~A.}\ \bibnamefont {Dill}},\ }\href
    {\doibase 10.1146/annurev-physchem-071119-040206} {\bibfield  {journal}
        {\bibinfo  {journal} {Annu. Rev. Phys. Chem.}\ }\textbf {\bibinfo {volume}
            {71}},\ \bibinfo {pages} {213} (\bibinfo {year} {2020})}\BibitemShut
    {NoStop}%
    \bibitem [{\citenamefont {Hedges}\ \emph {et~al.}(2009)\citenamefont {Hedges},
                \citenamefont {Jack}, \citenamefont {Garrahan},\ and\ \citenamefont
                {Chandler}}]{Hedges2009}%
    \BibitemOpen
    \bibfield  {author} {\bibinfo {author} {\bibfnamefont {L.~O.}\ \bibnamefont
            {Hedges}}, \bibinfo {author} {\bibfnamefont {R.~L.}\ \bibnamefont {Jack}},
        \bibinfo {author} {\bibfnamefont {J.~P.}\ \bibnamefont {Garrahan}}, \ and\
        \bibinfo {author} {\bibfnamefont {D.}~\bibnamefont {Chandler}},\ }\href
    {\doibase 10.1126/science.1166665} {\bibfield  {journal} {\bibinfo  {journal}
            {Science}\ }\textbf {\bibinfo {volume} {323}},\ \bibinfo {pages} {1309}
        (\bibinfo {year} {2009})}\BibitemShut {NoStop}%
    \bibitem [{\citenamefont {Chandler}\ and\ \citenamefont
                {Garrahan}(2010)}]{Chandler2010}%
    \BibitemOpen
    \bibfield  {author} {\bibinfo {author} {\bibfnamefont {D.}~\bibnamefont
            {Chandler}}\ and\ \bibinfo {author} {\bibfnamefont {J.~P.}\ \bibnamefont
            {Garrahan}},\ }\href {\doibase 10.1146/annurev.physchem.040808.090405}
    {\bibfield  {journal} {\bibinfo  {journal} {Annu. Rev. Phys. Chem.}\ }\textbf
        {\bibinfo {volume} {61}},\ \bibinfo {pages} {191} (\bibinfo {year}
        {2010})}\BibitemShut {NoStop}%
    \bibitem [{\citenamefont {Ye}\ \emph {et~al.}(2021)\citenamefont {Ye},
                \citenamefont {Zhuang},\ and\ \citenamefont {Li}}]{Ye2021}%
    \BibitemOpen
    \bibfield  {author} {\bibinfo {author} {\bibfnamefont {Q.-J.}\ \bibnamefont
            {Ye}}, \bibinfo {author} {\bibfnamefont {L.}~\bibnamefont {Zhuang}}, \ and\
        \bibinfo {author} {\bibfnamefont {X.-Z.}\ \bibnamefont {Li}},\ }\href
    {\doibase 10.1103/PhysRevLett.126.185501} {\bibfield  {journal} {\bibinfo
            {journal} {Phys. Rev. Lett.}\ }\textbf {\bibinfo {volume} {126}},\ \bibinfo
        {pages} {185501} (\bibinfo {year} {2021})}\BibitemShut {NoStop}%
    \bibitem [{\citenamefont {Geng}\ \emph {et~al.}(2017)\citenamefont {Geng},
                \citenamefont {Wu},\ and\ \citenamefont {Sun}}]{Geng2017}%
    \BibitemOpen
    \bibfield  {author} {\bibinfo {author} {\bibfnamefont {H.~Y.}\ \bibnamefont
            {Geng}}, \bibinfo {author} {\bibfnamefont {Q.}~\bibnamefont {Wu}}, \ and\
        \bibinfo {author} {\bibfnamefont {Y.}~\bibnamefont {Sun}},\ }\href {\doibase
        10.1021/acs.jpclett.6b02453} {\bibfield  {journal} {\bibinfo  {journal} {J.
                Phys. Chem. Lett.}\ }\textbf {\bibinfo {volume} {8}},\ \bibinfo {pages} {223}
        (\bibinfo {year} {2017})}\BibitemShut {NoStop}%
    \bibitem [{\citenamefont {Komatsu}\ \emph {et~al.}(2020)\citenamefont
                {Komatsu}, \citenamefont {Klotz}, \citenamefont {Machida}, \citenamefont
                {Sano-Furukawa}, \citenamefont {Hattori},\ and\ \citenamefont
                {Kagi}}]{Komatsu2020}%
    \BibitemOpen
    \bibfield  {author} {\bibinfo {author} {\bibfnamefont {K.}~\bibnamefont
            {Komatsu}}, \bibinfo {author} {\bibfnamefont {S.}~\bibnamefont {Klotz}},
        \bibinfo {author} {\bibfnamefont {S.}~\bibnamefont {Machida}}, \bibinfo
        {author} {\bibfnamefont {A.}~\bibnamefont {Sano-Furukawa}}, \bibinfo {author}
        {\bibfnamefont {T.}~\bibnamefont {Hattori}}, \ and\ \bibinfo {author}
        {\bibfnamefont {H.}~\bibnamefont {Kagi}},\ }\href {\doibase
        10.1073/pnas.1920447117} {\bibfield  {journal} {\bibinfo  {journal} {Proc.
                Natl. Acad. Sci. U. S. A.}\ }\textbf {\bibinfo {volume} {117}},\ \bibinfo
        {pages} {6356} (\bibinfo {year} {2020})}\BibitemShut {NoStop}%
    \bibitem [{\citenamefont {Wang}\ \emph {et~al.}(2021)\citenamefont {Wang},
                \citenamefont {Wang}, \citenamefont {Hermann}, \citenamefont {Liu},
                \citenamefont {Gao}, \citenamefont {Tosatti}, \citenamefont {Wang},
                \citenamefont {Xing},\ and\ \citenamefont {Sun}}]{Wang2021}%
    \BibitemOpen
    \bibfield  {author} {\bibinfo {author} {\bibfnamefont {Y.}~\bibnamefont
            {Wang}}, \bibinfo {author} {\bibfnamefont {J.}~\bibnamefont {Wang}}, \bibinfo
        {author} {\bibfnamefont {A.}~\bibnamefont {Hermann}}, \bibinfo {author}
        {\bibfnamefont {C.}~\bibnamefont {Liu}}, \bibinfo {author} {\bibfnamefont
            {H.}~\bibnamefont {Gao}}, \bibinfo {author} {\bibfnamefont {E.}~\bibnamefont
            {Tosatti}}, \bibinfo {author} {\bibfnamefont {H.-T.}\ \bibnamefont {Wang}},
        \bibinfo {author} {\bibfnamefont {D.}~\bibnamefont {Xing}}, \ and\ \bibinfo
        {author} {\bibfnamefont {J.}~\bibnamefont {Sun}},\ }\href {\doibase
        10.1103/PhysRevX.11.011006} {\bibfield  {journal} {\bibinfo  {journal} {Phys.
                Rev. X}\ }\textbf {\bibinfo {volume} {11}},\ \bibinfo {pages} {011006}
        (\bibinfo {year} {2021})}\BibitemShut {NoStop}%
    \bibitem [{\citenamefont {Queyroux}\ \emph {et~al.}(2020)\citenamefont
    {Queyroux}, \citenamefont {Hernandez}, \citenamefont {Weck}, \citenamefont
    {Ninet}, \citenamefont {Plisson}, \citenamefont {Klotz}, \citenamefont
    {Garbarino}, \citenamefont {Guignot}, \citenamefont {Mezouar}, \citenamefont
    {Hanfland}, \citenamefont {Iti{\'{e}}},\ and\ \citenamefont
    {Datchi}}]{Queyroux2020}%
    \BibitemOpen
    \bibfield  {author} {\bibinfo {author} {\bibfnamefont {J.-A.}\ \bibnamefont
        {Queyroux}}, \bibinfo {author} {\bibfnamefont {J.-A.}\ \bibnamefont
        {Hernandez}}, \bibinfo {author} {\bibfnamefont {G.}~\bibnamefont {Weck}},
    \bibinfo {author} {\bibfnamefont {S.}~\bibnamefont {Ninet}}, \bibinfo
    {author} {\bibfnamefont {T.}~\bibnamefont {Plisson}}, \bibinfo {author}
    {\bibfnamefont {S.}~\bibnamefont {Klotz}}, \bibinfo {author} {\bibfnamefont
        {G.}~\bibnamefont {Garbarino}}, \bibinfo {author} {\bibfnamefont
        {N.}~\bibnamefont {Guignot}}, \bibinfo {author} {\bibfnamefont
        {M.}~\bibnamefont {Mezouar}}, \bibinfo {author} {\bibfnamefont
        {M.}~\bibnamefont {Hanfland}}, \bibinfo {author} {\bibfnamefont {J.-P.}\
    \bibnamefont {Iti{\'{e}}}}, \ and\ \bibinfo {author} {\bibfnamefont
        {F.}~\bibnamefont {Datchi}},\ }\href {\doibase
        10.1103/PhysRevLett.125.195501} {\bibfield  {journal} {\bibinfo  {journal}
            {Phys. Rev. Lett.}\ }\textbf {\bibinfo {volume} {125}},\ \bibinfo {pages}
        {195501} (\bibinfo {year} {2020})}\BibitemShut {NoStop}%
    \bibitem [{\citenamefont {Lee}\ and\ \citenamefont {Yang}(1952)}]{Lee1952}%
    \BibitemOpen
    \bibfield  {author} {\bibinfo {author} {\bibfnamefont {T.~D.}\ \bibnamefont
            {Lee}}\ and\ \bibinfo {author} {\bibfnamefont {C.~N.}\ \bibnamefont {Yang}},\
    }\href {\doibase 10.1103/PhysRev.87.410} {\bibfield  {journal} {\bibinfo
            {journal} {Phys. Rev.}\ }\textbf {\bibinfo {volume} {87}},\ \bibinfo {pages}
        {410} (\bibinfo {year} {1952})}\BibitemShut {NoStop}%
    \bibitem [{\citenamefont {Fisher}(1967)}]{Fisher1967}%
    \BibitemOpen
    \bibfield  {author} {\bibinfo {author} {\bibfnamefont {M.~E.}\ \bibnamefont
            {Fisher}},\ }\href {\doibase 10.1088/0034-4885/30/2/306} {\bibfield
        {journal} {\bibinfo  {journal} {Rep. Prog. Phys.}\ }\textbf {\bibinfo
            {volume} {30}},\ \bibinfo {pages} {306} (\bibinfo {year} {1967})}\BibitemShut
    {NoStop}%
    \bibitem [{Note2()}]{Note2}%
    \BibitemOpen
    \bibinfo {note} {The form of the interacting potential might be more complex
        than the pairwise one. We adopt the pairwise for convenience of illustration
        and later computation, but the treatments are analogous for other forms of
        interacting potential.}\BibitemShut {Stop}%
    \bibitem [{SI()}]{SI}%
    \BibitemOpen
    \href@noop {} {}\bibinfo {note} {See supplemental material at http://xxx for
        details of the method and computational setups, as well as additional
        discussions, which further includes references {(\it 34-40)}.}\BibitemShut
    {Stop}%
    \bibitem [{\citenamefont {Yang}\ and\ \citenamefont {Lee}(1952)}]{Yang1952}%
    \BibitemOpen
    \bibfield  {author} {\bibinfo {author} {\bibfnamefont {C.~N.}\ \bibnamefont
            {Yang}}\ and\ \bibinfo {author} {\bibfnamefont {T.~D.}\ \bibnamefont {Lee}},\
    }\href {\doibase 10.1103/PhysRev.87.404} {\bibfield  {journal} {\bibinfo
            {journal} {Phys. Rev.}\ }\textbf {\bibinfo {volume} {87}},\ \bibinfo {pages}
        {404} (\bibinfo {year} {1952})}\BibitemShut {NoStop}%
    \bibitem [{\citenamefont {Fisher}(1965)}]{Fisher1965}%
    \BibitemOpen
    \bibfield  {author} {\bibinfo {author} {\bibfnamefont {M.}~\bibnamefont
            {Fisher}},\ }in\ \href
    {http://scholar.google.com/scholar?hl=en&btnG=Search&q=intitle:The+Nature+of+Critical+Points#0}
    {\emph {\bibinfo {booktitle} {Lectures in theoretical physics}}},\ Vol.\
    \bibinfo {volume} {VII},\ \bibinfo {editor} {edited by\ \bibinfo {editor}
        {\bibfnamefont {W.~E.}\ \bibnamefont {Britten}}}\ (\bibinfo  {publisher}
    {Univ. of Colorado Press, Boulder},\ \bibinfo {year} {1965})\ pp.\ \bibinfo
    {pages} {73--109}\BibitemShut {NoStop}%
    \bibitem [{\citenamefont {Wilczek}(2012)}]{Wilczek2012}%
    \BibitemOpen
    \bibfield  {author} {\bibinfo {author} {\bibfnamefont {F.}~\bibnamefont
            {Wilczek}},\ }\href {\doibase 10.1103/PhysRevLett.109.160401} {\bibfield
        {journal} {\bibinfo  {journal} {Phys. Rev. Lett.}\ }\textbf {\bibinfo
            {volume} {109}},\ \bibinfo {pages} {160401} (\bibinfo {year} {2012})}\
    \BibitemShut
    {NoStop}%
    \bibitem [{\citenamefont {Shapere}\ and\ \citenamefont
                {Wilczek}(2012)}]{Shapere2012}%
    \BibitemOpen
    \bibfield  {author} {\bibinfo {author} {\bibfnamefont {A.}~\bibnamefont
            {Shapere}}\ and\ \bibinfo {author} {\bibfnamefont {F.}~\bibnamefont
            {Wilczek}},\ }\href {\doibase 10.1103/PhysRevLett.109.160402} {\bibfield
        {journal} {\bibinfo  {journal} {Phys. Rev. Lett.}\ }\textbf {\bibinfo
            {volume} {109}},\ \bibinfo {pages} {160402} (\bibinfo {year}
        {2012})}\BibitemShut {NoStop}%
    \bibitem [{\citenamefont {Zhang}\ \emph {et~al.}(2017)\citenamefont {Zhang},
                \citenamefont {Hess}, \citenamefont {Kyprianidis}, \citenamefont {Becker},
                \citenamefont {Lee}, \citenamefont {Smith}, \citenamefont {Pagano},
                \citenamefont {Potirniche}, \citenamefont {Potter}, \citenamefont
                {Vishwanath}, \citenamefont {Yao},\ and\ \citenamefont {Monroe}}]{Zhang2017}%
    \BibitemOpen
    \bibfield  {author} {\bibinfo {author} {\bibfnamefont {J.}~\bibnamefont
            {Zhang}}, \bibinfo {author} {\bibfnamefont {P.~W.}\ \bibnamefont {Hess}},
        \bibinfo {author} {\bibfnamefont {A.}~\bibnamefont {Kyprianidis}}, \bibinfo
        {author} {\bibfnamefont {P.}~\bibnamefont {Becker}}, \bibinfo {author}
        {\bibfnamefont {A.}~\bibnamefont {Lee}}, \bibinfo {author} {\bibfnamefont
            {J.}~\bibnamefont {Smith}}, \bibinfo {author} {\bibfnamefont
            {G.}~\bibnamefont {Pagano}}, \bibinfo {author} {\bibfnamefont {I.~D.}\
            \bibnamefont {Potirniche}}, \bibinfo {author} {\bibfnamefont {A.~C.}\
            \bibnamefont {Potter}}, \bibinfo {author} {\bibfnamefont {A.}~\bibnamefont
            {Vishwanath}}, \bibinfo {author} {\bibfnamefont {N.~Y.}\ \bibnamefont {Yao}},
        \ and\ \bibinfo {author} {\bibfnamefont {C.}~\bibnamefont {Monroe}},\ }\href
    {\doibase 10.1038/nature21413} {\bibfield  {journal} {\bibinfo  {journal}
            {Nature}\ }\textbf {\bibinfo {volume} {543}},\ \bibinfo {pages} {217}
        (\bibinfo {year} {2017})}\BibitemShut {NoStop}%
    \bibitem [{\citenamefont {Heyl}\ \emph {et~al.}(2013)\citenamefont {Heyl},
                \citenamefont {Polkovnikov},\ and\ \citenamefont {Kehrein}}]{Heyl2013}%
    \BibitemOpen
    \bibfield  {author} {\bibinfo {author} {\bibfnamefont {M.}~\bibnamefont
            {Heyl}}, \bibinfo {author} {\bibfnamefont {A.}~\bibnamefont {Polkovnikov}}, \
        and\ \bibinfo {author} {\bibfnamefont {S.}~\bibnamefont {Kehrein}},\ }\href
    {\doibase 10.1103/PhysRevLett.110.135704} {\bibfield  {journal} {\bibinfo
            {journal} {Phys. Rev. Lett.}\ }\textbf {\bibinfo {volume} {110}},\ \bibinfo
        {pages} {135704} (\bibinfo {year} {2013})}\BibitemShut {NoStop}%
    \bibitem [{\citenamefont {Heyl}(2018)}]{Heyl2018}%
    \BibitemOpen
    \bibfield  {author} {\bibinfo {author} {\bibfnamefont {M.}~\bibnamefont
            {Heyl}},\ }\href {\doibase 10.1088/1361-6633/aaaf9a} {\bibfield  {journal}
        {\bibinfo  {journal} {Reports Prog. Phys.}\ }\textbf {\bibinfo {volume}
            {81}},\ \bibinfo {pages} {054001} (\bibinfo {year} {2018})}\BibitemShut
    {NoStop}%
    \bibitem [{\citenamefont {Smith}\ \emph {et~al.}(2016)\citenamefont {Smith},
                \citenamefont {Lee}, \citenamefont {Richerme}, \citenamefont {Neyenhuis},
                \citenamefont {Hess}, \citenamefont {Hauke}, \citenamefont {Heyl},
                \citenamefont {Huse},\ and\ \citenamefont {Monroe}}]{Smith2016}%
    \BibitemOpen
    \bibfield  {author} {\bibinfo {author} {\bibfnamefont {J.}~\bibnamefont
            {Smith}}, \bibinfo {author} {\bibfnamefont {A.}~\bibnamefont {Lee}}, \bibinfo
        {author} {\bibfnamefont {P.}~\bibnamefont {Richerme}}, \bibinfo {author}
        {\bibfnamefont {B.}~\bibnamefont {Neyenhuis}}, \bibinfo {author}
        {\bibfnamefont {P.~W.}\ \bibnamefont {Hess}}, \bibinfo {author}
        {\bibfnamefont {P.}~\bibnamefont {Hauke}}, \bibinfo {author} {\bibfnamefont
            {M.}~\bibnamefont {Heyl}}, \bibinfo {author} {\bibfnamefont {D.~A.}\
            \bibnamefont {Huse}}, \ and\ \bibinfo {author} {\bibfnamefont
            {C.}~\bibnamefont {Monroe}},\ }\href {\doibase 10.1038/nphys3783} {\bibfield
        {journal} {\bibinfo  {journal} {Nat. Phys.}\ }\textbf {\bibinfo {volume}
            {12}},\ \bibinfo {pages} {907} (\bibinfo {year} {2016})}\BibitemShut
    {NoStop}%
    \bibitem [{\citenamefont {Zhuang}\ \emph {et~al.}(2020)\citenamefont {Zhuang},
                \citenamefont {Ye}, \citenamefont {Pan},\ and\ \citenamefont
                {Li}}]{Zhuang2020}%
    \BibitemOpen
    \bibfield  {author} {\bibinfo {author} {\bibfnamefont {L.}~\bibnamefont
            {Zhuang}}, \bibinfo {author} {\bibfnamefont {Q.-J.}\ \bibnamefont {Ye}},
        \bibinfo {author} {\bibfnamefont {D.}~\bibnamefont {Pan}}, \ and\ \bibinfo
        {author} {\bibfnamefont {X.-Z.}\ \bibnamefont {Li}},\ }\href {\doibase
        10.1088/0256-307X/37/4/043101} {\bibfield  {journal} {\bibinfo  {journal}
            {Chinese Phys. Lett.}\ }\textbf {\bibinfo {volume} {37}},\ \bibinfo {pages}
        {043101} (\bibinfo {year} {2020})}\BibitemShut {NoStop}%
    \bibitem [{\citenamefont {Kresse}\ and\ \citenamefont
    {Furthm{\"{u}}ller}(1996)}]{Kresse1996}%
    \BibitemOpen
    \bibfield  {author} {\bibinfo {author} {\bibfnamefont {G.}~\bibnamefont
        {Kresse}}\ and\ \bibinfo {author} {\bibfnamefont {J.}~\bibnamefont
    {Furthm{\"{u}}ller}},\ }\href {\doibase 10.1103/PhysRevB.54.11169} {\bibfield
        {journal} {\bibinfo  {journal} {Phys. Rev. B}\ }\textbf {\bibinfo {volume}
            {54}},\ \bibinfo {pages} {11169} (\bibinfo {year} {1996})}\BibitemShut
    {NoStop}%
    \bibitem [{\citenamefont {Kresse}\ and\ \citenamefont
                {Joubert}(1999)}]{Kresse1999}%
    \BibitemOpen
    \bibfield  {author} {\bibinfo {author} {\bibfnamefont {G.}~\bibnamefont
            {Kresse}}\ and\ \bibinfo {author} {\bibfnamefont {D.}~\bibnamefont
            {Joubert}},\ }\href {\doibase 10.1103/PhysRevB.59.1758} {\bibfield  {journal}
        {\bibinfo  {journal} {Phys. Rev. B}\ }\textbf {\bibinfo {volume} {59}},\
        \bibinfo {pages} {1758} (\bibinfo {year} {1999})}\BibitemShut {NoStop}%
    \bibitem [{\citenamefont {Sun}\ \emph {et~al.}(2015)\citenamefont {Sun},
                \citenamefont {Ruzsinszky},\ and\ \citenamefont {Perdew}}]{Sun2015}%
    \BibitemOpen
    \bibfield  {author} {\bibinfo {author} {\bibfnamefont {J.}~\bibnamefont
            {Sun}}, \bibinfo {author} {\bibfnamefont {A.}~\bibnamefont {Ruzsinszky}}, \
        and\ \bibinfo {author} {\bibfnamefont {J.~P.}\ \bibnamefont {Perdew}},\
    }\href {\doibase 10.1103/PhysRevLett.115.036402} {\bibfield  {journal}
        {\bibinfo  {journal} {Phys. Rev. Lett.}\ }\textbf {\bibinfo {volume} {115}},\
        \bibinfo {pages} {036402} (\bibinfo {year} {2015})}\BibitemShut {NoStop}%
    \bibitem [{\citenamefont {Wang}\ \emph {et~al.}(2018)\citenamefont {Wang},
                \citenamefont {Zhang}, \citenamefont {Han},\ and\ \citenamefont
                {E}}]{Wang2018}%
    \BibitemOpen
    \bibfield  {author} {\bibinfo {author} {\bibfnamefont {H.}~\bibnamefont
            {Wang}}, \bibinfo {author} {\bibfnamefont {L.}~\bibnamefont {Zhang}},
        \bibinfo {author} {\bibfnamefont {J.}~\bibnamefont {Han}}, \ and\ \bibinfo
        {author} {\bibfnamefont {W.}~\bibnamefont {E}},\ }\href {\doibase
        10.1016/j.cpc.2018.03.016} {\bibfield  {journal} {\bibinfo  {journal}
            {Comput. Phys. Commun.}\ }\textbf {\bibinfo {volume} {228}},\ \bibinfo
        {pages} {178} (\bibinfo {year} {2018})}\BibitemShut {NoStop}%
    \bibitem [{\citenamefont {Plimpton}(1995)}]{PLIMPTON19951}%
    \BibitemOpen
    \bibfield  {author} {\bibinfo {author} {\bibfnamefont {S.}~\bibnamefont
            {Plimpton}},\ }\href {\doibase 10.1006/jcph.1995.1039} {\bibfield  {journal}
        {\bibinfo  {journal} {J. Comput. Phys.}\ }\textbf {\bibinfo {volume} {117}},\
        \bibinfo {pages} {1} (\bibinfo {year} {1995})}\BibitemShut {NoStop}%
    \bibitem [{\citenamefont {Zoller}\ \emph {et~al.}(1987)\citenamefont {Zoller},
                \citenamefont {Marte},\ and\ \citenamefont {Walls}}]{Zoller1987}%
    \BibitemOpen
    \bibfield  {author} {\bibinfo {author} {\bibfnamefont {P.}~\bibnamefont
            {Zoller}}, \bibinfo {author} {\bibfnamefont {M.}~\bibnamefont {Marte}}, \
        and\ \bibinfo {author} {\bibfnamefont {D.~F.}\ \bibnamefont {Walls}},\ }\href
    {\doibase 10.1103/PhysRevA.35.198} {\bibfield  {journal} {\bibinfo  {journal}
            {Phys. Rev. A}\ }\textbf {\bibinfo {volume} {35}},\ \bibinfo {pages} {198}
        (\bibinfo {year} {1987})}\BibitemShut {NoStop}%
\end{thebibliography}
%

\end{document}